\documentclass[aps,pra,twocolumn,superscriptaddress,showpacs]{revtex4-2}

\usepackage[utf8]{inputenc}

\usepackage{array}
\usepackage{amsmath}
\usepackage{graphicx}
\usepackage{multirow}
\usepackage{epstopdf}
\usepackage{color}
\usepackage{amsfonts}
\usepackage{subfigure}
\usepackage{amssymb}
\usepackage{graphicx}
\usepackage{makecell}
\usepackage[colorlinks,citecolor=blue]{hyperref}
\usepackage{changes}
\newcommand{\onecolm}{
  \end{multicols}
  \vspace{-3.5ex}
  \noindent\rule{0.5\textwidth}{0.1ex}\rule{0.1ex}{2ex}\hfill
}
\newcommand{\twocolm}{
  \hfill\raisebox{-1.9ex}{\rule{0.1ex}{2ex}}\rule{0.5\textwidth}{0.1ex}
  \vspace{-4ex}
  \begin{multicols}{2}
}

\begin{document}

\title{Survival of Hermitian Criticality in the Non-Hermitian Framework}
\author{Fei Wang}
\affiliation{Institute of Biophysics, Dezhou University, Dezhou 253023, China}
\author{Guoying Liang}
\affiliation{Institute of Biophysics, Dezhou University, Dezhou 253023, China}
\author{Zecheng Zhao}
\affiliation{Institute of Biophysics, Dezhou University, Dezhou 253023, China}
\author{Lin-Yue Luo}
\affiliation {Key Laboratory of Atomic and Subatomic Structure and Quantum Control (Ministry of Education), Guangdong Basic Research Center of Excellence for Structure and Fundamental Interactions of Matter, School of Physics, South China Normal University, Guangzhou 510006, China}
\affiliation {Guangdong Provincial Key Laboratory of Quantum Engineering and Quantum Materials, Guangdong-Hong Kong Joint Laboratory of Quantum Matter, Frontier Research Institute for Physics, South China Normal University, Guangzhou 510006, China}
\author{Da-Jian Zhang}
\email{zdj@sdu.edu.cn}
\affiliation {Department of Physics, Shandong University, Jinan 250100, China}
\author{Bao-Ming Xu}
\email{xubm2018@163.com}
\affiliation{Institute of Biophysics, Dezhou University, Dezhou 253023, China}
\date{Submitted \today}

\begin{abstract}
In this work, we investigate many-body phase transitions in a one-dimensional anisotropic XY model subject to a complex-valued transverse field. Within the biorthogonal framework, we calculate the ground-state correlation functions and entanglement entropy, confirming that their scaling behavior remains identical to that in the Hermitian XY model. The preservation of Hermitian phase transition features in the non-Hermitian setting is rooted in the persistence and emergence of symmetries and their breaking. Specifically, the ferromagnetic (FM) phase arises from the breaking of a $Z_2$ symmetry, while the Luttinger liquid (LL) phase is enabled by the emergence of a $U(1)$ symmetry together with the degeneracy of the real part of the energy spectrum. The nontrivial topology of the LL phase is characterized by the winding number around the exceptional point (EP). Given that non-Hermitian systems are inherently open, our findings suggest a new avenue for exploring universal critical phenomena associated with conventional quantum phase transitions, which are typically vulnerable to decoherence and environmental disruption in conventional open quantum systems.
\end{abstract}

\maketitle

\section{Introduction}\label{Intr}
Quantum phase transitions have long been a central research theme in quantum many-body physics. Notable examples include quantum magnet-paramagnet transitions \cite{Lieb1961,Barouch1971}, Mott insulator-superfluid transitions \cite{Fisher1989,Lee2006}, superconductor-insulator transitions in two-dimensional systems \cite{Liu1994}, and topological quantum phase transitions \cite{Hasan2010,Moore2010,Qi2011,Li2008,Pollmann2010}. These transitions take place at absolute zero temperature when an external parameter or coupling constant is tuned through a critical point, driven by quantum fluctuations \cite{Sachdev2000}. A quantum phase is characterized by the long-range properties of the system, such as the spin-spin correlation function \cite{Barouch1971}. Furthermore, quantum entanglement, as a purely quantum effect, has been widely used to characterize quantum phase transitions \cite{Amico2008,Eisert2010,Osterloh2002,Vidal2003}. Entanglement entropy scaling laws are, for example, able to discriminate between different universality classes of gapless phases, in particular in one dimension \cite{Vidal2003,Calabrese2004,Korepin2004,Francesco1996}, but also can include terms that have a topological origin and characterize the fundamental topological excitations of the system \cite{Kitaev2006,Levin2006}. At the quantum critical point between distinct phases, physical quantities exhibit singular behavior reflecting changes in the long-range properties, governed by the symmetries of the system Hamiltonian \cite{Vojta2003}.

Typical quantum phase transitions are generally studied in terms of Hermitian Hamiltonians. However, realistic experimental platforms, such as cold-atom systems \cite{Diehl2008,Li2020a}, optical cavities \cite{Ozturk2021,Baumann2010}, waveguides \cite{Ke2016,Sun2022a}, and optomechanical devices \cite{Pino2022,Wang2024}, are inherently non-Hermitian due to spontaneous decay. This discrepancy raises a fundamental question: Whether quantum phase transitions and critical phenomena originally identified in Hermitian systems can persist in their non-Hermitian counterparts? If so, through what mechanisms is this criticality manifested in non-Hermitian systems? Typical Quantum phase transitions, especially in low dimensions, are highly sensitive to thermal fluctuations and often require near-zero temperatures for a sharp manifestation. If such transitions can be recovered in non-Hermitian systems, it would imply that the critical behavior (e.g., critical exponents and scaling relations) may remain robust against certain types of engineered non-Hermitian perturbations. This robustness suggests that the universal aspects of the phase transition are encoded in a way that transcends the Hermitian constraint. Consequently, highly controllable non-Hermitian platforms can serve as simulators, where tuning designed gain/loss parameters allows one to probe and access the universal physics of the corresponding Hermitian quantum critical point, even at finite temperatures. Advances in non-Hermitian physics have opened up new avenues for exploring novel types of phase transition specific to non-Hermitian systems \cite{Ashida2020,Ganainy2018,Okuma2023,Bergholtz2021,Ding2022}. However, few studies have investigated -- from the perspective of traditional quantum phase transitions in Hermitian systems -- whether the intrinsic critical behavior of such systems can persist or be meaningfully expressed within non-Hermitian frameworks.

To address this issue, this work considers a one-dimensional anisotropic XY model subject to a complex valued transverse field to study the quantum phase transitions and critical phenomena in its ground state. Unlike Hermitian systems, which possess real eigenvalues and orthogonal eigenstates, general non-Hermitian Hamiltonians may have complex eigenvalues and non-orthogonal eigenstates. To be more precise, the orthogonality of eigenstates is replaced by the notion of biorthogonality that defines the relation between the right and left eigenvectors, leading to the so-called ``biorthogonal quantum mechanics" \cite{Brody2014}. The biorthogonal theoretical framework is recognized as correctly capturing the nature of states in non-Hermitian systems. As a result, it has been extensively adopted to investigate phase transitions in non-Hermitian quantum systems \cite{Jing2024,Edvardsson2020,Zhang2025,Sun2022b,Lu2024,Liu2025,Deng2025,Lu2025b,Lu2025c,Longhi2023,Para2021,Hu2021,Lin2022,Zhang2024,Couvreur2017,Deng2024,Solinas2025,Yang2020b,Lee2020}. In this framework, we calculate the longitudinal spin-spin correlation function and entanglement entropy. Our results show that quantum phase transitions -- typically susceptible to decoherence and environmental disruption in Hermitian systems -- can be robustly sustained in their non-Hermitian analogues. Remarkably, the critical scaling properties of the correlation function and entanglement entropy in such non-Hermitian settings remain exactly the same as those observed in Hermitian systems. To uncover the underlying mechanism behind the preservation of Hermitian phase transition features in the non-Hermitian setting, we analyze the energy spectrum of the non-Hermitian XY model. Our analysis reveals that phase transitions are governed by the persistence and emergence of symmetries and their spontaneous breaking in the ground state. Specifically, the ferromagnetic (FM) phase originates from the breaking of a $Z_2$ symmetry, while the LL phase is enabled by the emergence of a $U(1)$ symmetry, accompanied by the degeneracy of the real part of the energy spectrum. Furthermore, to characterize the nontrivial topology of the LL phase, we introduce a winding number defined with respect to the exceptional point (EP). This topological invariant effectively captures the anomalous geometric structure associated with the LL phase under non-Hermitian conditions. It must be pointed out that the biorthogonal theoretical framework plays a crucial role in the survival of traditional Hermitian phase transitions in non-Hermitian systems. Conversely, if one considers only the right eigenvector, the observation of the Hermitian phase transition becomes impossible.

This paper is organized as follows: In Sec. \ref{NH XY Model}, we introduce the non-Hermitian one-dimensional XY model and the biorthogonal theoretical framework. Sec. \ref{PT and Cri} presents a study of the phase transitions in non-Hermitian XY model. In Sec. \ref{Underlying M}, the underlying mechanism behind the preservation of Hermitian phase transition features in the non-Hermitian setting is discussed. Finally, Sec. \ref{Conclusions} closes the paper with some concluding remarks.

\section{Non-Hermitian XY model}\label{NH XY Model}
We consider the non-Hermitian one-dimensional XY model whose Hamiltonian is
\begin{equation}\label{}
  \hat{H}=-\frac{J}{2}\sum_{j=1}^{N}\biggl[\frac{1+\gamma}{2}\hat{\sigma}_j^x\hat{\sigma}_{j+1}^x
  +\frac{1-\gamma}{2}\hat{\sigma}_j^y\hat{\sigma}_{j+1}^y\biggr]+\frac{\lambda}{2}\sum_{j=1}^N\hat{\sigma}_j^z,
\end{equation}
where $\hat{\sigma}^{\alpha}_{j}$ $(\alpha=x,y,z)$ is the spin-1/2 Pauli operator at lattice site $j$ and the periodic boundary condition is imposed as $\hat{\sigma}^{\alpha}_{N+1}=\hat{\sigma}^{\alpha}_{1}$. Here we only consider that $N$ is even. $J$ is the longitudinal spin-spin coupling, and we set $J=1$ as the overall energy scale without loss of generality. $\gamma$ governs the anisotropic coupling between spins along the $x$ and $y$ directions, and $\lambda$ is a dimensionless parameter measuring the strength of the transverse field with respect to the longitudinal spin-spin coupling. We consider the transverse field is complex $\lambda=\lambda_0e^{i\phi}$, where $\lambda_0$ and $\phi$ are the absolute value and argument of $\lambda$, respectively.

\begin{figure}
\begin{center}
\includegraphics[width=8cm]{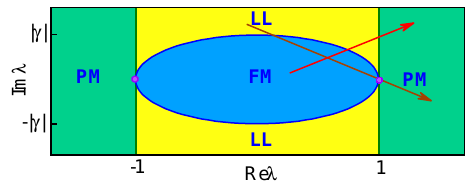}
\caption{(Color online) The phase diagram of the non-Hermitian XY model as a function of $\mathrm{Re}\lambda$ and $\mathrm{Im}\lambda$ for a given $\gamma$. The blue elliptical interior ($(\mathrm{Re}\lambda)^2+(\mathrm{Im}\lambda)^2/\gamma^2<1$) corresponds to the FM phase. The green region ($|\mathrm{Re}\lambda| > 1$) represents the PM phase. The rest yellow region [$(\mathrm{Re}\lambda)^2+(\mathrm{Im}\lambda)^2/\gamma^2>1$ and $\mathrm{Re}\lambda < 1$] exhibits the glass phase. The purple circles represent the critical point corresponding to FM-PM phase transition, i.e., $\lambda=1$, in the Hermitian XY model.}
\label{fig1}
\end{center}
\end{figure}

The Hamiltonian is integrable and can be mapped to a system of free fermions and therefore be solved exactly. By applying the Jordan-Wigner transformation and the Fourier transformation, the Hamiltonain converts from spin operators into spinless fermionic operators as \cite{Lieb1961}
\begin{equation}\label{Hamiltonian}
\hat{H}=\sum_{k>0}
\begin{pmatrix} \hat{c}^\dag_{k} & \hat{c}_{-k} \end{pmatrix}
\begin{pmatrix} \lambda-\cos k &-i\gamma\sin k \\ i\gamma\sin k & -\lambda+\cos k\end{pmatrix}
\begin{pmatrix} \hat{c}_{k} \\ \hat{c}^\dag_{-k} \end{pmatrix},
\end{equation}
where $\hat{c}_{k}$ and $\hat{c}^\dag_{k}$ are respectively fermion annihilation and creation operators for mode $k=(2n-1)\pi/N$ with $n=1,2,\cdots, N/2$, corresponding to antiperiodic boundary conditions when $N$ is even. Its ground state energy is
\begin{equation}\label{ground energy}
 \varepsilon_k^-=-\sqrt{(\lambda-\cos k)^2+(\gamma\sin k)^2}
\end{equation}
with the corresponding right and left eigenvectors
\begin{equation}\label{}
\begin{split}
|g_{k}\rangle&=\biggl[\cos\frac{\theta_k}{2}+i\sin\frac{\theta_k}{2}\hat{c}^\dag_{k}\hat{c}^\dag_{-k}\biggr]|0\rangle, \\
\langle \tilde{g}_{k}|&=\langle 0|\biggl[\cos\frac{\theta_k}{2}-i\sin\frac{\theta_k}{2}\hat{c}_{-k}\hat{c}_{k}\biggr]
\end{split}
\end{equation}
satisfying $\hat{H}_k|g_{k}\rangle=\varepsilon^-_k|g_{k}\rangle$, $\langle \tilde{g}_{k}|\hat{H}_k=\langle \tilde{g}_{k}|\varepsilon^-_k$ and $\langle \tilde{g}_{k}|g_{k}\rangle=1$. $|0\rangle$ is the vacuum state. The angle $\theta_k$, which is generally complex, is determined by $\tan\theta_k=\gamma\sin k/(\lambda-\cos k)$.

\begin{figure*}
\centering
\includegraphics[width=16cm]{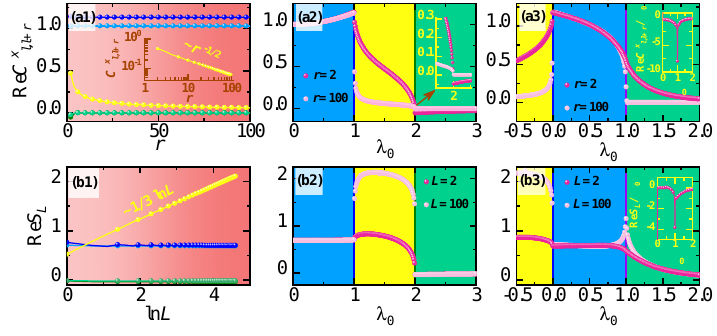}
\caption{(Color online) (a1)-(b1) The longitudinal spin-spin correlation function $\mathrm{Re}C_{l,l+r}^x$ and the entanglement entropy $\mathrm{Re}S_L$ as a function of spin-spin distance $r$ and the subsystem size $L$ for $\gamma=1$ and $\lambda=\lambda_0e^{i\pi/3}$ with $\lambda_0=0.5$ (bright blue), $\lambda_0=1$ (blue), $\lambda_0=1.5$ (yellow), $\lambda_0=2$ (olive) and $\lambda_0=3$ (light green). The inset of (a1) shows the scaling law of the spin-spin correlation at $\lambda=1.5e^{i\pi/3}$. (a2)-(b2) The longitudinal spin-spin correlation function $\mathrm{Re}C_{l,l+r}^x$ and the entanglement entropy $\mathrm{Re}S_L$ as functions of $\lambda_0$ for $\gamma=1$ and $\lambda=\lambda_0e^{i\pi/3}$ (along the red arrow in Fig. \ref{fig1}). (a3)-(b3) The longitudinal spin-spin correlation function $\mathrm{Re}C_{l,l+r}^x$ and the entanglement entropy $\mathrm{Re}S_L$ as functions of $\lambda_0$ for $\gamma=1$ and $\lambda=\lambda_0+i(1-\lambda_0)$ (along the brown arrow in Fig. \ref{fig1}). The insets of (a3) and (b3) display the derivation of the short spin-spin correlation $\mathrm{Re}C_{l,l+2}^x$ and local entanglement $\mathrm{Re}S_2$, respectively.}
\label{fig2}
\end{figure*}

\section{Phase transition and criticality}\label{PT and Cri}
The Hermitian XY model with real valued $\lambda$ exhibits competitions between anisotropic and magnetic couplings, which results in the existence of multiple phases. The quantum phase transition from the ferromagnetic phase (FM) to the paramagnetic phase (PM) driven by the transverse field $\lambda$ is called the Ising transition with the quantum critical point $\lambda = 1$. On the other hand, the quantum phase transition between two FMs, with magnetic ordering in the $x$-direction and the $y$-direction, respectively, driven by the anisotropy parameter $\gamma$, is called the anisotropic transition with the critical point $\gamma = 0$. In fact, when $\lambda<1$, the ground state of the system at $\gamma = 0$ is in the Luttinger liquid (LL) phase \cite{Dutta2010}. The main goal of this paper is to examine whether these quantum phase transitions and critical phenomena can persist in their non-Hermitian counterparts? If so, through what mechanisms are these critical properties manifested in non-Hermitian systems?

A great deal more information about phase transition may be obtained by investigating the longitudinal spin-spin correlation function \cite{Lieb1961}
\begin{equation}\label{}
  C_{ll+r}^x=\mathrm{Tr}[\hat{\rho}_g\hat{\sigma}_l^x\hat{\sigma}_{l+r}^x],
\end{equation}
where $\hat{\rho}_g=\bigotimes_k|g_{k}\rangle\langle \tilde{g}_{k}|$ is ground-state density matrix. In addition, entanglement is appointed to play a central role in understanding quantum phase transitions \cite{Vidal2003}. The entanglement entropy is defined as
\begin{equation}\label{}
  S_L=-\mathrm{Tr}[\hat{\rho}_L\ln\hat{\rho}_L],
\end{equation}
where $\hat{\rho}_L=\mathrm{Tr}_{\bar{L}}[\hat{\rho}_g]$ is the reduced density matrix of a bock of $L$ spins, obtained by tracing out the degrees of freedom in the rest of the system. Detailed derivations of these quantities within the biorthogonal theory framework are provided in the Appendix. The results demonstrate that precisely due to the adoption of the biorthogonal theoretical framework, the expressions for the correlation function and entanglement entropy in the non-Hermitian system closely resemble those in the Hermitian case. Taking the Hermitian limit naturally recovers the expressions of the Hermitian system. This strong formal similarity provides a basis for preserving the phase transition and critical behavior of Hermitian systems within non-Hermitian settings. In the non-Hermitian regime, both spin-spin correlation functions and entanglement entropy become complex-valued; we therefore probe phase transitions exclusively through their real parts.

From Eq. \eqref{ground energy}, we can see that the energy becomes degenerate when $\lambda=\cos k\pm i\gamma\sin k$, corresponding to a set of EPs. All EPs lie on an ellipse (Fig. \ref{fig1}) satisfying
\begin{equation}\label{}
(\mathrm{Re}\lambda)^2+\frac{(\mathrm{Im}\lambda)^2}{\gamma^2}=1,
\end{equation}
whose shape depends on the anisotropy parameter $\gamma$. Within the elliptical region, the system exhibits strong spin-spin correlations, with $\mathrm{Re}C_{l,l+r}^r$ remaining close to 1 regardless of the distance $r$ [bright blue curve, Fig. \ref{fig2}(a1)]. Furthermore, it sustains finite entanglement entropy $\mathrm{Re}S_L$ that is independent of subsystem size $L$ [bright blue curve, Fig. \ref{fig2}(b1)]. These properties indicate robust long-range order persistence across arbitrary spatial scales, characterizing the FM phase. Conversely, for $\mathrm{Re}\lambda>1$, both spin-spin correlations and entanglement entropy vanish [green curves, Fig. \ref{fig2}(a1, b1)], confirming a disordered PM phase. Crucially, the interplay of power-law decay in the spin-spin correlation $\mathrm{Re}C_{l,l+r}^x\sim r^{-1/2}$ and logarithmic scaling of the entanglement entropy $\mathrm{Re}S_L\sim1/3\ln L$ [yellow curves, Fig. \ref{fig2} (a1, b1)] in the rest region defined by $(\mathrm{Re}\lambda)^2+\frac{(\mathrm{Im}\lambda)^2}{\gamma^2}>1$ and $\mathrm{Re}\lambda < 1$ identifies a LL phase. The LL phase eludes conventional long-range order, as its correlation functions decay algebraically rather than saturating to a constant. However, it sustains a profound, distinct long-range organization rooted in quantum entanglement. This is quantified by the logarithmic scaling of its entanglement entropy. This entanglement structure acts as a non-local order parameter, defining the criticality, scale-invariant nature beyond the Landau symmetry-breaking paradigm. A non-Hermitian XY model with a complex field was previously studied by considering only the right eigenvectors in Ref. \cite{Liu2021}. Within that formulation, their calculated correlation function exhibits a combined form of power-law and exponential decay, which deviates from the pure power-law behavior characteristic of a critical phase. In contrast, our work is grounded in the biorthogonal framework, which allows us to identify the universality class of the critical phase explicitly as a LL phase.

Phase transitions induced by transverse field variations along different paths (arrows in Fig. \ref{fig1}) are shown in Fig. \ref{fig2} (a2-a3), (b2-b3). It is noteworthy that the LL-PM and LL-FM phase transitions trigger abrupt changes in correlation functions and entanglement entropy at their respective critical points. In contrast, the correlation functions and entanglement entropy exhibit continuous variation across the PM-FM phase transition. At the LL-FM critical points, indicated by the blue vertical lines in Fig. \ref{fig2}(a2-a3) and (b2-b3), the correlation and entanglement properties match those of the FM phase. Similarly, at the LL-PM critical points where $\mathrm{Re}\lambda = 1$, marked by the green vertical lines in Fig. \ref{fig2}(a2-a3) and (b2-b3), these properties align with those of the PM phase.

Remarkably, these scaling behaviors and the phase transition information are identical to those of Hermitian XY model. This discovery demonstrates that the quantum phase transitions and critical behavior inherent to the Hermitian system can be preserved within the non-Hermitian framework, affirming the remarkable resilience of Hermitian universality classes within a non-Hermitian context.

\section{Underlying mechanisms} \label{Underlying M}
A central question is how phase transitions, well-understood in Hermitian systems, carry over to non-Hermitian ones. To address this, we must first recall the underpinnings of the Hermitian paradigm, where (broken) symmetry plays a crucial role. The Hermitian anisotropic XY model possesses a global $Z_2$ symmetry, remaining invariant under the transformations $\sigma_j^x\rightarrow-\sigma_j^x$, $\sigma_j^y\rightarrow-\sigma_j^y$ and $\sigma_j^z\rightarrow\sigma_j^z$. This discrete symmetry is preserved due to the energy gap between the unique ground state and excited states in the PM phase ($\lambda>1$), whereas it is spontaneously broken at the critical point $\lambda = 1$ as the system collapses into one of two degenerate ground states, marking a conventional symmetry-breaking phase transition (Ising transition). Beyond this, the XY model turns to the XX model at $\gamma=0$, manifesting an inherent $U(1)$ symmetry. Although this symmetry remains unbroken in the ground state, it fundamentally constrains the phase behavior of the system by prohibiting the opening of an energy gap in the critical regime ($\lambda\leq1$). The closure of the energy gap at the quantum critical point leads to the emergence of protected Fermi points in momentum space, which dictate a specific occupation pattern where only states below the Fermi energy are occupied. This selective occupation configuration induces a change in the topological invariant, compelling the system to reside in a gapless LL phase instead of a trivial insulator.

\begin{figure}
\begin{center}
\includegraphics[width=8cm]{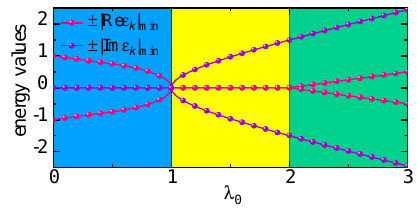}
\caption{(Color online) The minimum values of the real and imaginary parts of the energy as the function of $\lambda_0$ with $\lambda=\lambda_0e^{i\pi/3}$ and $\gamma=1$.}
\label{fig3}
\end{center}
\end{figure}

Now we turn our attention to the non-Hermitian XY model. The non-Hermitian XY model still preserves $Z_2$ symmetry. At the EPs defined by $(\mathrm{Re}\lambda)^2 + \frac{(\mathrm{Im}\lambda)^2}{\gamma^2} = 1$, the $Z_2$ symmetry is spontaneously broken in the ground state as the energy gap closes, giving rise to the FM phases within the region $(\mathrm{Re}\lambda)^2 + \frac{(\mathrm{Im}\lambda)^2}{\gamma^2} < 1$. Although $Z_2$ symmetry remains unbroken for $(\mathrm{Re}\lambda)^2 + \frac{(\mathrm{Im}\lambda)^2}{\gamma^2} > 1$, the system enters the PM phase only when $\mathrm{Re}\lambda > 1$; in the rest of this regime, it exhibits LL behavior.

To reveal the underlying mechanism of the LL phase in the non-Hermitian XY model, we focus our attention on the energy-level structure of the system. In Fig. \ref{fig3}, we plot the minimal values of the real and imaginary parts of the energy, defined as  $|\mathrm{Re}\varepsilon_k|_{min}=\min_k|\mathrm{Re}\varepsilon_k|$ and $|\mathrm{Im}\varepsilon_k|_{min}=\min_k|\mathrm{Im}\varepsilon_k|$, respectively. It should be pointed out that the mode $k$ associated with the minimal real part of the energy generally differs from that corresponding to the minimal imaginary part. A striking phase dependence is observed in the energy-level structure. Within the FM phase, the imaginary part of the energy can become degenerate, whereas the real part does not. Conversely, in the LL phase, degeneracy is possible in the real part but not in the imaginary part. The PM phase exhibits no degeneracy in either component. In the non-Hermitian context, the degeneracy of the real part of the energy accounts for the emergence of the LL phase as the closure of the energy gap does in its Hermitian counterpart.

\begin{figure}
\begin{center}
\includegraphics[width=8cm]{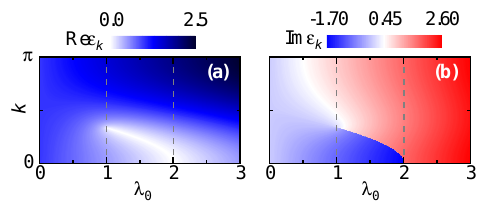}
\caption{(Color online) The real (a) and imaginary (b) parts of the full energy across all mode as the function of $\lambda_0$. The parameters are $\lambda=\lambda_0e^{i\pi/3}$ and $\gamma=1$.}
\label{fig4}
\end{center}
\end{figure}

From Eqs. \eqref{Hamiltonian} and \eqref{ground energy}, it can be seen that the degeneracy of the real part of the energy requires $\mathrm{Re}\lambda-\cos k=0$ and $|\mathrm{Im}\lambda|>|\gamma\sin k|$, which determine the critical mode $k_c=\arccos\mathrm{Re}\lambda$ with a purely imaginary energy. Surprisingly, we find that the ground state associated with this critical mode, $\hat{\rho}_{k_c}=|g_{k_{c}}\rangle\langle \tilde{g}_{k_{c}}|$, satisfies the relation
\begin{equation}\label{}
  R_z(\theta)\Bigl(\hat{\rho}_{k_c}+\hat{\rho}_{k_c}^\dag\Bigr)R_z^\dag(\theta)=\hat{\rho}_{k_c}+\hat{\rho}_{k_c}^\dag
\end{equation}
for any $\theta$, where $R_z(\theta)=\exp(-i\theta\sum_j\sigma_j^z)=\exp(-i\theta\sum_kc_k^\dag c_k)$ is the $U(1)$ rotation operator around the $z$-axis. This $U(1)$ symmetry is manifested in the real part of a physical observation, such as the real parts of the correlation function $\mathrm{Re}C_{l,l+r}^x$ and the entanglement entropy $\mathrm{Re}S_L$ considered in this paper. It is important to emphasize that neither the system Hamiltonian nor its ground state possesses $U(1)$ symmetry--rather, the $U(1)$ symmetry emerges spontaneously in the critical mode where the real part of the energy is degenerate. This emergent $U(1)$ symmetry underlies the LL scaling behavior observed in the correlation function and entanglement entropy in Fig. \ref{fig2}.

\begin{table}[]
    \centering
    \caption{Comparison of phase transition conditions between the Hermitian and non-Hermitian XY models.}
    \label{tab:phase_comparison}
    \renewcommand{\arraystretch}{1.5} 
    \begin{tabular}{lcc}
        \hline
        & \textbf{Hermitian} & \textbf{Non-Hermitian} \\
        \hline\hline
        \multirow{2}{*}{\textbf{FM}}~ & $|\lambda|<1$, $\gamma\neq0$
                                    & $(\mathrm{Re}\lambda)^2+\frac{(\mathrm{Im}\lambda)^2}{\gamma^2}<1$   \\
                                    & Broken $Z_2$ symmetry~   & Broken $Z_2$ symmetry                     \\
        \hline
        \multirow{2}{*}{\textbf{PM}}~ & $|\lambda|>1$              & $|\mathrm{Re}\lambda|>1$   \\
                                    & $Z_2$ symmetry             & $Z_2$ symmetry              \\
        \hline
        \multirow{3}{*}{\textbf{LL}}~ & $|\lambda|<1$, $\gamma=0$
                                    & $(\mathrm{Re}\lambda)^2+\frac{(\mathrm{Im}\lambda)^2}{\gamma^2}>1$, $|\mathrm{Re}\lambda|<1$ \\
                                    & $U(1)$ symmetry            & Emergent U(1) symmetry           \\
                                    & Gapless energy                & Gapless real-energy          \\
        \hline
    \end{tabular}
\end{table}

\begin{figure}
\begin{center}
\includegraphics[width=8cm]{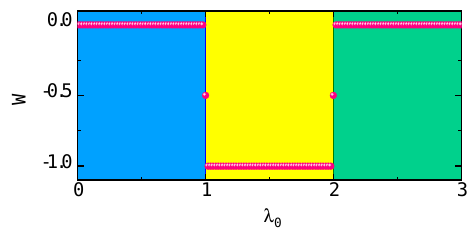}
\caption{(Color online) The winding number for different phases as the function of $\lambda_0$ for $\lambda=\lambda_0e^{i\pi/3}$ and $\gamma=1$.}
\label{fig5}
\end{center}
\end{figure}

Furthermore, a striking sign reversal in the imaginary part of the energy is observed across the full spectrum of modes [Fig. \ref{fig4}(b)]. Remarkably, this sign reversal occurs precisely at the degeneracy point of the real part of the energy [Fig. \ref{fig4}(a)]. A negative imaginary part indicates that the corresponding state is unstable and will eventually dissipate, implying that only the mode with a positive imaginary part becomes occupied. This phenomenon, where a selective population of a specific mode governs the low-energy physics of the system, bears a functional analogy to the LL phase in Hermitian systems. In both cases, the selective occupation dictates a change in the topological properties. To describe these topological properties in non-Hermitian setting, we define a topological invariant:
\begin{equation}\label{winding number}
W = \frac{1}{2\pi i} \int_0^{2\pi} \frac{\partial}{\partial k} \arg\mathbf{r}(k)dk,
\end{equation}
where $\mathbf{r}(k) = \mathbf{d}(k) - \mathbf{d}_e(k_c)$ with $\mathbf{d}(k) \equiv d_z(k) + i d_y(k) = \lambda - \cos k + i \gamma \sin k$ being a complex vector derived from the Hamiltonian $\hat{H}_k = \mathbf{d}(k) \cdot \bm{\sigma}$. $\mathbf{d}_e(k_c) = \lambda_e - \cos k_c + i \gamma \sin k_c$ is the reference vector at the EP $\lambda_e$ with critical momentum $k_c = \arccos(\mathrm{Re}\lambda_e)$. EP $\lambda_e$ is determined by $(\mathrm{Re}\lambda)^2 + \frac{(\mathrm{Im}\lambda)^2}{\gamma^2} = 1$. Thus, the topological invariant in Eq.~\eqref{winding number} can be interpreted as the winding number around this EP. In Fig.~\ref{fig5}, we plot the winding number across different phases. A jump to $W = -1$ within the LL phase reveals nontrivial topological character, while at the phase boundaries, the winding number takes the value $-1/2$. The comparison of phase transition conditions between the Hermitian and non-Hermitian XY models are summarized in Table. \ref{tab:phase_comparison}.

Finally, it is essential to emphasize that the phase transition behavior in the non-Hermitian system can only be properly captured within the biorthogonal framework. If one considers only the right eigenvector of the Hamiltonian \cite{Liu2021}, the results described above cannot be recovered. This can be understood as follows: a non-Hermitian Hamiltonian can not be fully expanded using only right eigenvectors, i.e., $\hat{H}\neq\sum_nE_n|\psi_n\rangle\langle\psi_n|$ with $\hat{H}|\psi_n\rangle=E_n|\psi_n\rangle$; a complete description requires both left and right eigenvectors, i.e., $\hat{H}=\sum_nE_n|\psi_n\rangle\langle\tilde{\psi}_n|$ where $\hat{H}|\psi_n\rangle=E_n|\psi_n\rangle$ and $\langle\tilde{\psi}_n|\hat{H}=\langle\tilde{\psi}_n|E_n$. In other words, the right eigenvector alone does not encode all the physical information of the system. Only within the biorthogonal formalism -- which incorporates both left and right eigenstates -- can the underlying symmetry of the non-Hermitian Hamiltonian and its breaking be consistently reflected.

\section{Conclusions and outlook} \label{Conclusions}
In this paper, we have investigated the quantum phase transitions and critical phenomena in the ground state of the one-dimensional anisotropic XY model with a complex-valued transverse field, using biorthogonal theory to examine longitudinal spin-spin correlation function and entanglement entropy. Notably, we found that the conventional quantum phase transitions in Hermitian systems can be faithfully retained in their non-Hermitian counterparts. The preservation of Hermitian phase transition features in the non-Hermitian setting is rooted in the persistence and emergence of symmetries and their breaking. Specifically, the FM phase arises from the breaking of a $Z_2$ symmetry, while the LL phase is enabled by the degeneracy of the real part of the energy spectrum together with the emergence of a $U(1)$ symmetry. To further characterize the LL phase, we introduced a winding number defined around EPs, revealing its nontrivial topological nature.

The survival of Hermitian phase transitions in non-Hermitian regimes is of particular significance. While such transitions are generally susceptible to environmental decoherence in Hermitian systems, their stability in inherently open non-Hermitian contexts offers a promising platform for exploring quantum critical phenomena under realistic conditions. This insight opens a new avenue for studying conventional phase transitions in open quantum systems, where dissipation and decoherence can no longer be overlooked.

\section*{Acknowledgments}
D.-J. Zhang acknowledges support from the National Natural Science Foundation of China (Grant Nos. 12275155 and 12174224).

\begin{appendix}
\section{The diagonalization of non-Hermitian XY model}
The Hamiltonian of the non-Hermitian one-dimensional XY model is integrable and can be mapped to a system of free fermions by applying the Jordan-Wigner transformation \cite{Lieb1961}:
\begin{equation}\label{}
  \begin{split}
    \hat{\sigma}_j^+ & =\exp\Bigl[-i\pi\sum_{j'=1}^{j-1}\hat{c}_{j'}^\dag \hat{c}_{j'}\Bigr]\hat{c}_j \\
    \hat{\sigma}_j^- & =\hat{c}_j^\dag\exp\Bigl[i\pi\sum_{j'=1}^{j-1}\hat{c}_{j'}^\dag \hat{c}_{j'}\Bigr],
  \end{split}
\end{equation}
satisfying the anticommutation relation
\begin{equation}\label{}
\begin{split}
  \{\hat{c}_{j},\hat{c}_{j'}^\dag\}=\delta_{jj'}, \\
  \{\hat{c}_{j},\hat{c}_{j'}\}=\{\hat{c}_{j}^\dag,\hat{c}_{j'}^\dag\}=0.
\end{split}
\end{equation}
It should be noted that
\begin{equation}\label{}
  \exp\Bigl[-i\pi\hat{c}_{j}^\dag \hat{c}_{j}\Bigr]=\exp\Bigl[i\pi\hat{c}_{j}^\dag \hat{c}_{j}\Bigr]=1-2\hat{c}_{j}^\dag \hat{c}_{j}=\hat{\sigma}_j^z.
\end{equation}
After the Jordan-Wigner transformation, the Hamiltonian of the system can be expressed as
\begin{equation}\label{HJW}
  \hat{H}=-\frac{J}{2}\sum_{j}\biggl[\hat{c}_{j}^\dag\hat{c}_{j+1}
  +\gamma\hat{c}_{j}^\dag\hat{c}_{j+1}^\dag+\lambda\hat{c}_{j}^\dag\hat{c}_{j}+h.c.\biggr],
\end{equation}
where we have considered the antiperiodic boundary conditions when $N$ is even and neglected the correction term related to the parity.
Consider the Fourier transformation
\begin{equation}\label{}
  \begin{split}
    \hat{c}_j & =\frac{1}{\sqrt{N}}\hat{c}_ke^{ijk}, \\
    \hat{c}_j^\dag & =\frac{1}{\sqrt{N}}\hat{c}_k^\dag e^{-ijk},
  \end{split}
\end{equation}
Eq. \eqref{HJW} can be expressed as
\begin{equation}\label{Hamiltonian}
\hat{H}=\sum_{k>0}
\begin{pmatrix} \hat{c}^\dag_{k} & \hat{c}_{-k} \end{pmatrix}
\begin{pmatrix} \lambda-\cos k &-i\gamma\sin k \\ i\gamma\sin k & -\lambda+\cos k\end{pmatrix}
\begin{pmatrix} \hat{c}_{k} \\ \hat{c}^\dag_{-k} \end{pmatrix},
\end{equation}
where $\hat{c}_{k}$ and $\hat{c}^\dag_{k}$ are respectively fermion annihilation and creation operators for mode $k=(2n-1)\pi/N$ with $n=1,2,\cdots, N/2$.
\section{Correlation function}
The longitudinal spin-spin correlation function is defined as
\begin{equation}\label{}
  C_{ll+r}^x=\mathrm{Tr}[\hat{\rho}_g\hat{\sigma}_l^x\hat{\sigma}_{l+r}^x],
\end{equation}
In terms of Jordan-Wigner fermions
\begin{equation}\label{}
  C_{ll+r}^x=\mathrm{Tr}\Biggl[\hat{\rho}_g(\hat{c}_l^\dag+\hat{c}_l)\exp\Biggl(i\pi\sum_{j=l}^{l+r-1}\hat{c}_j^\dag \hat{c}_j\Biggr)(\hat{c}_{l+r}^\dag+\hat{c}_{l+r})\Biggr].
\end{equation}
In the representation diagonalizing $\hat{c}_j^\dag \hat{c}_j$, it can be verified that
\begin{equation}\label{}
  \exp(i\pi \hat{c}_j^\dag \hat{c}_j)=(\hat{c}_j^\dag+c_j)(\hat{c}_j^\dag-c_j).
\end{equation}
Defining $\hat{A}_j=\hat{c}_j^\dag+\hat{c}_j$ and $\hat{B}_j=\hat{c}_j^\dag-\hat{c}_j$, we have
\begin{equation}\label{}
  C_{ll+r}^x=\mathrm{Tr}\bigl[\hat{\rho}_g\hat{B}_l\hat{A}_{l+1}\hat{B}_{l+1}\cdots \hat{A}_{l+r-1}\hat{B}_{l+r-1}\hat{A}_{l+r}\bigr].
\end{equation}
The complicated expression can be simplified using Wick's theorem. We find that in the biorthogonal theory framework,
\begin{equation}\label{}
\begin{split}
  &\mathrm{Tr}\bigl[\hat{\rho}_g\hat{A}_{j}\hat{A}_{j'}\bigr]=\delta_{jj'},\\
  &\mathrm{Tr}\bigl[\hat{\rho}_g\hat{B}_{j}\hat{B}_{j'}\bigr]=-\delta_{jj'}.
  \end{split}
\end{equation}
And only nonzero contractions are $\mathrm{Tr}[\hat{\rho}_g\hat{A}_{j}\hat{B}_{j'}]$ and $\mathrm{Tr}\bigl[\hat{\rho}_g\hat{B}_{j}\hat{A}_{j'}\bigr]$.
After the Fourier transformation,
\begin{equation}\label{}
  \mathrm{Tr}[\hat{\rho}_g\hat{A}_{j}\hat{B}_{j+r}]=\frac{2}{N}\sum_{k>0}\Bigl[\sin\theta_k\sin(rk)+\cos\theta_k\cos(rk)\Bigr]
\end{equation}
and
\begin{equation}\label{}
  \mathrm{Tr}[\hat{\rho}_g\hat{B}_{j}\hat{A}_{j+r}]=\frac{2}{N}\sum_{k>0}\Bigl[\sin\theta_k\sin(rk)-\cos\theta_k\cos(rk)\Bigr].
\end{equation}
In the thermodynamic limit $N\rightarrow\infty$, they can be expressed as
\begin{equation}\label{}
  \mathrm{Tr}[\hat{\rho}_g\hat{A}_{j}\hat{B}_{j+r}]=\frac{1}{\pi}\int_0^\pi dk\Bigl[\sin\theta_k\sin(rk)+\cos\theta_k\cos(rk)\Bigr]
\end{equation}
and
\begin{equation}\label{}
  \mathrm{Tr}[\hat{\rho}_g\hat{B}_{j}\hat{A}_{j+r}]=\frac{1}{\pi}\int_0^\pi dk\Bigl[\sin\theta_k\sin(rk)-\cos\theta_k\cos(rk)\Bigr].
\end{equation}

Defining $\mathrm{Tr}[\hat{\rho}_g\hat{B}_{j}\hat{A}_{j+r}]=-\mathrm{Tr}[\hat{\rho}_g\hat{A}_{j+r}\hat{B}_{j}]=G_r$, the correlation function is given by a determinant
\begin{equation}\label{}
  C_{ll+r}^x=
  \begin{vmatrix}
  & G_{1}    &  G_{2}      &  \cdots      &  G_{r}      \\
  & G_{0}    &  G_{1}      &  \cdots      &  G_{r-1}    \\
  & \vdots   & \vdots      &  \ddots      &  \vdots     \\
  & G_{-r+2} & G_{-r+1}    &  \cdots      &  G_{1}
  \end{vmatrix}.
\end{equation}

\section{Entanglement Entropy}
In Hermitian mechanics, the entanglement of the state $|\Psi_{ab}\rangle$ of systems $a$ and $b$ can be measured by the entanglement entropy $E(|\Psi_{ab}\rangle)=-\mathrm{Tr}[\hat{\rho}_a\ln\hat{\rho}_a]$, where $\hat{\rho}_a=\mathrm{Tr}_b[|\Psi_{ab}\rangle\langle\Psi_{ab}|]$. Here, alternatively, we undertake the study of the entanglement in the non-Hermitian systems. Specifically, we divide the spin chain into two parts: a block containing $L$ spins and a block containing the remaining spins, and investigate the quantum entanglement between these two blocks. For the ground state $\hat{\rho}_g$, the entanglement between two blocks is defined as
\begin{equation}\label{}
  S_L=-\mathrm{Tr}[\hat{\rho}_L\ln\hat{\rho}_L],
\end{equation}
where $\hat{\rho}_L=\mathrm{Tr}_{\bar{L}}[\hat{\rho}_g]$ is the reduced density matrix of the bock of $L$ spins. $\mathrm{Tr}_{\overline{L}}$ denotes tracing out the degrees of freedom of the rest spins other than block $L$. It should be noted that $S_L$ is complex because $\hat{\rho}_L$ is non-Hermitian.

In order to calculate entanglement entropy, we need to define, for each site $l$ of the $N$-spin chain, two Majorana operators \cite{Vidal2003}
\begin{equation}\label{}
  \check{c}_{2l}=\Biggl(\prod_{m<l}\hat{\sigma}_m^z\Biggr)\hat{\sigma}_l^x,
  ~~\check{c}_{2l+1}=\Biggl(\prod_{m<l}\hat{\sigma}_m^z\Biggr)\hat{\sigma}_l^y
\end{equation}
Operators $\check{c}_m$ are Hermitian and obey the anticommutation relations $\{\check{c}_m,\check{c}_n\}=2\delta_{mn}$. The expectation value of $\check{c}_m$ when the system is in the ground state, i.e., $\langle\check{c}_m\rangle=\mathrm{Tr}[\hat{\rho}_g\check{c}_m]$, vanishes for all $m$ due to the $Z_2$ symmetry $(\hat{\sigma}_l^x,\hat{\sigma}_l^y,\hat{\sigma}_l^z)\rightarrow(-\hat{\sigma}_l^x,-\hat{\sigma}_l^y,\hat{\sigma}_l^z)$ of the original Hamiltonian. Majorana operators can be expressed by Fermionic creation and annihilation operators
\begin{equation}\label{}
  \check{c}_{2l}=\hat{c}_l^\dag+\hat{c}_l=\hat{A}_l,
  ~~\check{c}_{2l+1}=-i(\hat{c}_l^\dag-\hat{c}_l)=-i\hat{B}_l.
\end{equation}

In turn, the expectation values $\langle \check{c}_m\check{c}_n\rangle=\delta_{mn}+i\Gamma_{mn}$ completely characterize $\hat{\rho}_g$, for any other expectation value can be expressed, through Wick's theorem, in terms of $\langle \check{c}_m\check{c}_n\rangle$. Matrix $\Gamma$ reads
\begin{equation}\label{GammaMatr}
  \Gamma=
  \begin{pmatrix}
  & \Pi_0       & \Pi_1      &  \cdots      & \Pi_{N-1}  \\
  & \Pi_{-1}      & \Pi_0      &  \cdots      & \vdots     \\
  & \vdots      & \vdots     &  \ddots      &  \vdots  \\
  & \Pi_{1-N}   & \cdots     &  \cdots      & \Pi_0
  \end{pmatrix},
\end{equation}
where
\begin{equation}\label{}
    \Pi_r=-i
    \begin{pmatrix}
  & \langle \check{c}_{2l}\check{c}_{2(l+r)}\rangle           & \langle \check{c}_{2l}\check{c}_{2(l+r)+1}\rangle    \\
  & \langle \check{c}_{2l+1}\check{c}_{2(l+r)}\rangle     & \langle \check{c}_{2l+1}\check{c}_{2(l+r)+1}\rangle      \\
  \end{pmatrix}.
\end{equation}
Using Wick's theorem $\Pi_r$ can be calculated as
\begin{equation}\label{}
    \Pi_r=
  \begin{pmatrix}
  & 0           & G_{-r}    \\
  & -G_r     & 0      \\
  \end{pmatrix}.
\end{equation}
Eliminating the rows and columns corresponding to those spins of the chain that do not belong to the block $L$, we compute the correlation matrix of the state $\hat{\rho}_L$, namely $\delta_{mn}+i(\Gamma_L)_{mn}$, where
\begin{equation}\label{GammaMatrL}
  \Gamma=
  \begin{pmatrix}
  & \Pi_0       & \Pi_1      &  \cdots      & \Pi_{L-1}  \\
  & \Pi_{-1}      & \Pi_0      &  \cdots      & \vdots     \\
  & \vdots      & \vdots     &  \ddots      & \vdots     \\
  & \Pi_{1-L}   & \cdots     &  \cdots      & \Pi_0
  \end{pmatrix}.
\end{equation}
Diagonalize $\Gamma_L$ we can obtain its eigenvalues $\pm\nu_m$ ($m=0,1,\cdots L-1$), and thus the entanglement entropy is
\begin{equation}\label{}
  S_L=\sum_{m=0}^{L-1}H_2\Bigl(\frac{1+\nu_m}{2}\Bigr)
\end{equation}
with $H_2(x)=-x\ln x-(1-x)\ln(1-x)$.

Notably, within the biorthogonal theoretical framework, the expressions for correlation function and entanglement entropy bear a strong formal resemblance to those in the Hermitian XY model, apart from their complex-valued nature stemming from non-Hermiticity. In the Hermitian limit, these expressions naturally reduce to their Hermitian counterparts. This strong formal similarity provides an analytical foundation for preserving the phase transition and critical behavior of Hermitian systems within non-Hermitian settings.

\end{appendix}

\end{document}